\documentclass[aps,prb,twocolumn,amssymb,amsmath]{revtex4}
\usepackage[dvips]{graphicx}
\begin{document}

\title{%
Transport in Nanotubes: Effect of Remote Impurity Scattering
}

\author{%
Alexey G. Petrov$^{1,2}$, Slava V. Rotkin$^{1,2}$ }

%\address
\affiliation{(1) Beckman Institute, UIUC, 405 N.Mathews, Urbana,
IL 61801, USA;\\
(2) Ioffe Institute, 26 Politekhnicheskaya st.,
St.Petersburg 194021, Russia.
\\ E--mail: rotkin@uiuc.edu
}

\begin{abstract}
Theory of the remote Coulomb impurity scattering in single--wall
carbon nanotubes is developed within one--electron
approximation. Boltzmann equation is solved within
drift--diffusion model to obtain the tube conductivity. The
conductivity depends on the type of the nanotube bandstructure
(metal or semiconductor) and on the electron Fermi level. We
found exponential dependence of the conductivity on the Fermi
energy due to the Coulomb scattering rate has a strong
dependence on the momentum transfer. We calculate intra-- and
inter--subband scattering rates and present general expressions
for the conductivity. Numerical results, as well as obtained
analytical expressions, show that the degenerately doped
semiconductor tubes may have very high mobility unless the
doping level becomes too high and the inter--subband transitions
impede the electron transport.
\end{abstract}
\maketitle

\section{Introduction \label{sec:intro}}

Carbon nanotubes, discovered last decade\cite{iijima}, attract
attention of physicists not only due to their beautiful
symmetry, extreme quantum scale and quasi--one--dimensional (1D)
structure, but also due to numerous application, already
existing\cite{avrev,av2,cathods} and foreseen in future. Among
those, the applications, which will use transport properties of
nanotubes\cite{av2,appenz}, are sensitive to details of a charge
carrier scattering. It was shown that the scattering in an ideal
{\em metallic} single--wall nanotube (SWNT) is greatly
diminished as in a full symmetry of an armchair nanotube the
backscattering between states at a Fermi level (in a thin shell
near an electrochemical potential at $T\neq 0$) is
forbidden\cite{ftnt}. In contrast, an electron in a doped {\em
semiconductor} SWNT, which has different symmetry (zigzag or
other chiral symmetry), can be backscattered. In field effect
devices the semiconductor nanotubes are used. To obtain high
conductivity the doping level is controlled, thus, our study of
the transport mechanisms and the semiconductor SWNT conductivity
\cite{footnote1} and their dependence on the doping level
becomes very timely.

Among most important mechanisms of elastic \cite{andorev}
scattering, responsible for the low temperature resistivity of
nanoscale systems, one may encounter a defect scattering, a
Coulomb center scattering, and an electron--electron
interaction; the latter is beyond the scope of one--electron
theory and will be studied elsewhere. First two are distinct as
a scattering potential is long ranged for the Coulomb center,
and it is short ranged for a neutral (mechanical) defect, which
may be a substitutional (not charged) impurity or another type
of a lattice defect. The charge carrier scattering by the
Coulomb center, charged impurity and/or ionized dopant is in the
focus of our paper. At low temperatures this mechanism is
comparable or even stronger than other scattering mechanisms.
Modern high mobility semiconductor devices have so small size
scale that the device channel may contain no substitutional
impurity/dopant atom. The remote impurity scattering may still
limit the transport in this case. Recent experiments
\cite{fuhrer,johnson} showed that even long channel
semiconductor SWNT devices may have very high mobility at the
high doping level. Our theory addresses this case, for the first
time taking into account Coulomb scattering at the remote
impurities and including effects of an inter--subband scattering
at the high (doping) Fermi level.

Specifics of the Coulomb interaction in nanotubes is that the
effective potential, seen by the delocalized electrons, has a cut
off at the radius of the tube, $R$. Thus, even if the Coulomb
impurity is at the closest (atomic) distance from the tube surface
\cite{footnote2}, the effective Coulomb potential is the potential
of a remote scatterer. The situation is similar to what is known
for HEMT devices. We propose the remote impurity scattering
mechanism to be responsible for a residual resistance of the
semiconductor nanotube at low temperature. It may define a limit
for an electron mobility (ON current) in the nanotube field effect
transistor if other scattering mechanisms are less effective. Same
theory is applicable to the scattering in metallic nanotubes which
are proposed for use as interconnects and ballistic wires
\cite{meyya} and especially in metallic field effect transistors
\cite{metfet}.

We develop below a theory of the remote impurity scattering for
the SWNT device, calculate a scattering rate at one surface
impurity, scattering by a dilute distribution of random
impurities (in a Born approximation), a zero and finite
temperature conductivity in a drift--diffusion (DD) model. It is
known that in pure (mesoscopic) systems, a quantum interference
correction to the DD conductivity depends on the system
dimension and at $D=1$ it is of the order of the conductivity
itself. This manifests as a 1D localization. However, for a
quasi--1D system of the nanotube it may not hold due to fast
phase breaking. It may explain why considerably simple classical
one--electron theory describes the nanotube experiment fairly
well \cite{ando,tersoff,shik}. Among possible mechanisms
responsible for the fast phase breaking we notice: e--e and
e--phonon interactions. The latter mechanism was shown to be
very fast in SWNTs in optical studies \cite{hertel}. We will
discuss the role of e--e interaction later on.

We calculate dependence of the SWNT conductivity on the Fermi
level. Strong degeneracy of carriers in the semiconductor tubes
is required to obtain the high mobility \cite{johnson2}. We show
that this is consistent with calculated scattering rates. The
same condition of high Fermi level allows one to apply the
Boltzmann equation for transport calculations.

It is commonly used that the conduction in SWNTs happens via
states of one subband (of orbital quantization). We note that at
the high doping level the new conduction channel (through the
next subband) must appear. Then, an interesting question opens:
how the scattering between {\em different subbands} may change
the conductivity. In semiconductors role of an intervalley
scattering is not very important. The difference in the
transport at high fields consists mainly in the renormalization
of an effective mass of the electron of the conduction band. We
will show that for the quasi--1D bandstructure of the SWNT,
where a phase space of scattering is very restricted, the
opening of a new scattering channel may change the conductivity
qualitatively, especially for the metallic armchair nanotubes
and for high mobility semiconducting nanotubes. We found that
for the Coulomb scattering mechanism it results in a giant drop
of the conductivity at low temperature. We present a general
expression for the concentration dependence of the conductivity
and study its analytical limits that clearly demonstrates the
physics of this effect.

1D conductivity, $\sigma_{1D}$, is known to have a dimension of
a diffusion coefficient,
$\langle$length$\rangle^2/\langle$time$\rangle$ (see for
example, Ref. \cite{furman}). Thus, the conductance of the 1D
system is $\sigma_{1D}/L$ where $L$ is the system length. We
demonstrate below that a simple expression is valid for the DD
conductivity of the SWNT: $\sigma_{1D}\sim G_o \Lambda$, where
$G_o=e^2/2\pi\hbar$ is the quantum of conductance and $\Lambda
\sim v \tau$ is a mean free path of the electron ($v$ and $\tau$
are the electron velocity and lifetime). We calculated $\Lambda$
as a function of the Fermi energy and a strength of the random
impurity potential. Within the model discussed, the conductance
can be written as:
\begin{equation}
G \sim \sigma_{1D} \frac{1}{L} \sim G_o \frac{R}{L}
f_1\left(\frac{E_F}{U_s}\right) \,
f_2\left(\frac{E_F}{E_g}\right) \, g\left(\frac{E_F}{E_g}\right)
\label{appr}
\end{equation}
here $R$ is the SWNT radius (characteristic transverse length of
the 1D channel), $E_F$ is the electron Fermi level, $U_s$ is a
strength of the 1D potential of the Coulomb impurities, $E_g$ is
the SWNT gap, $f_1$ and  $f_2$ are some power law functions, and
$g(x)$ is an exponential function of its argument. The
conductivity (and conductance) is so sensitive to the change of
the Fermi level (exponential function of $E_F$) because of the
exponential dependence of $g(x)$ that reflects a strong
dependence of the Coulomb matrix element on a transferred
momentum. Allowed backscattering transitions within a single
subband have a large momentum transfer at large $E_F$, while the
transition between subbands may have smaller momentum transfer.
Thus, an appearing of the new scattering channel drastically
decreases the mean free path of the electron, $\Lambda$, which
becomes small and $G$ drops several orders of magnitude. We
notice that the power law function $f_1(x)\gg 1$ because of its
argument is large: $E_F\gg U_s$. This inequality is also a
necessary condition for applicability of the Boltzmann equation.

Below we consider charged impurities which are likely presented
on a surface of any substrate. The deposition methods which are
currently used in fabrication of the nanotube devices may
produce such impurities in large quantities. Charged impurities
are known to exist at the surface of SiO$_2$ substrate, commonly
used for nanotube devices. Use of high--$\epsilon$ dielectrics
may even increase the role of this scattering mechanism.

We assume that the impurities are (single) ionized and uniformly
distributed on the insulating surface with a 2D density $n_s$.
Both assumptions are not vital for the model and make no
qualitative change in the final results. However, the derivation
of the Eq.(\ref{appr}), for example, is more clear in this
geometry. The generalization of our theory to the case of 3D
distribution of the Coulomb centers is straightforward. It gives
a description for the SWNT embedded in an insulating matrix and
will be published elsewhere \cite{unpubl}. For the Coulomb
substitutional impurities located directly in the lattice of the
SWNT and for the charged impurities encapsulated inside the
tube, an order of magnitude estimate for the DD conductivity can
be obtained by substituting $U_s^2$ by $W_s E_c$, where $W_s$ is
the strength of a random potential of 1D impurities and $E_c\sim
e^2/C$ is a Coulomb charging energy of the tube. We notice the
logarithmic divergence of the latter with the $L$ as well as the
same logarithmic divergence of the 1D Coulomb matrix
element\cite{loudon}.

\section{Remote impurity scattering rate \label{sec:scat}}
\subsection{Model assumptions \label{sec:assumpt}}

The nanotube is situated at the van der Waals distance from the
surface of the substrate. This distance is about 3.4~\AA, and the
Coulomb centers are removed from the device channel. The Born
approximation, which implies independent scattering events, is
used in what follows to calculate the elastic scattering rate. We
consider scattering of the electrons in different subbands and
between subbands. One can apply this theory to the transport in
multiwall nanotubes, which are believed to conduct by the
outermost shell. Here, we restrict ourselves to the case of
single--wall tubes and consider armchair and zigzag SWNTs, though,
the final result is more general and can be, possibly, used for an
arbitrary tube.

The Coulomb scattering manifests itself in low--dimensional
systems. The long range Coulomb potential is known to be
underscreened in a 1D case \cite{loudon,oned}, in nanowires
\cite{shik} and nanotubes \cite{rider,ieee}. Thus, the Coulomb
scattering becomes the most important scattering mechanism at
certain conditions. It is well known that the random potential
results in a localization of carriers in an infinite 1D
system\cite{and}. We assume that an effective phase breaking
mechanism exists in the SWNTs, which destroys the interference.
It is known that the e--e scattering time is very short for
these systems. Even though, the e--e interaction in {\em a
single 1D band} cannot suppress the localization. Because the
total momentum of the electronic subsystem is conserved.
However, in the nanotubes there is a number of different
e--channels which may be not coherent. It is clear that the e--e
scattering between the electrons that belong to {\em the
different bands} breaks the phase of the wave function and,
thus, destroy the interference. What are these different
e--bands? The transport in nanotubes occurs via the band formed
by highest valence electrons, so called, $\pi$--electron band.
The rest of the valence electrons are localized and form
low--lying $\sigma$--bands.  The e--e scattering between these
two different bands does not preserve the electron phase.  Also,
in a real experimental situation, in the SWNT rope or in the
multiwall nanotube, the scattering between $\pi$--electrons at
the different walls/tubes destroys the interference as well. For
what follows we accept that the phase breaking time is short
enough to neglect the interference correction and use the
Boltzmann equation approach for the calculation of the DD
conductivity of the SWNT.

\subsection{Envelope wave functions\label{sec:model}}

We use SWNT envelope wave functions \cite{meleold} for the
electronic structure calculation, which are obtained as a
solution of a tight--binding (TB) Hamiltonian for $\pi$
electrons \cite{dress}. Our approach is very close to what was
presented in Ref.\cite{meleold} and then widely used in the
nanotube literature, so we skip details and give only the final
wave functions for the two--band scheme ($\pi$ electrons only):
\begin{equation}
|\psi_{m,k,\zeta}\rangle=\frac{1}{\sqrt{2}}(|A\rangle +\zeta
c_{mk}|B\rangle)
e^{ikz} e^{im\alpha},
\label{wavefun}
\end{equation}
here an orbital momentum $m$ labels orbital subbands of the SWNT
electronic structure, $k$ labels states with a longitudinal
momentum, both $m$ and $k$ are good quantum numbers (discrete
and continuum, respectively) for an ideal, long enough nanotube.
$\zeta=\pm 1$ is a pseudospin. A pseudospinor vector is formed
by a two--component amplitude of the wave function defined for
two atoms in a graphite unit cell (A and B). Coordinate along
the tube is $z$, and $\alpha$ is an angle along the nanotube
circumference (by this we explicitly assume the electron to be
confined to a surface of a cylinder of fixed radius $R$). The
components of the pseudospinor are c--numbers, in general,
defined up to an arbitrary phase. It is taken such that the
coefficient for the A--component is purely real and equal to
$1/\sqrt{2}$, then a matrix element of the dimensionless TB
Hamiltonian gives the second component of the pseudospinor
\cite{gonzalez}. This determines a dependence of $c_{mk}$ on the
subband index, $m$, and the 1D momentum, $k$. The pseudospin
$\zeta$ distinguishes between states of valence and conduction
band \cite{damnjan}.

So far, we considered an equilibrium (non--perturbed) electronic
wave functions. We assume, as usual, that the scattering can be
modelled perturbatively if the interference terms are negligible
as it was discussed in the last section. The perturbation operator
is the Coulomb potential:
\begin{equation}
\displaystyle V_i({\bf r})=
\frac{e e^*}{\sqrt{(z-Z_i)^2+(x-X_i)^2+(y-Y_i)^2}},
\label{potential}
\end{equation}
where $e<0$ is an electron charge. $x$, $y$ and $z$ are the
coordinates of the electron. These three coordinates are not
independent as the electron motion is restricted to the surface
of the cylinder. $e^*$ is an effective charge of an impurity,
its position is given in a Cartesian coordinate system as
$[X_i,Y_i,Z_i]$. For the scattering at a single impurity, a
relative position of the Coulomb center along the (infinite)
nanotube, $Z_i$, may be chosen arbitrary. The coordinate $X_i$
(normal to the substrate surface) approximately equals $h$, a
negative height of the nanotube, which assumes that the impurity
size is negligible and that the impurity is not buried in the
substrate (both assumptions are reasonable but the model works
without this simplification as well). We will define an
effective charge of the impurity, $e^*$, in the last section,
when discussing the screening.

\subsection{Matrix element of the impurity potential \label{sec:me}}

One needs to know matrix elements of the Coulomb potential
between the TB wave functions of the electron to calculate the
scattering. The potential of the remote impurity is smooth at
the surface of the nanotube and, therefore, it is almost
constant within the unit cell. Hence, the matrix elements
of the potential with the envelope wave functions
(\ref{wavefun})
can be approximated by the
1D Fourier components of Eq.(\ref{potential}):
\begin{eqnarray}
\begin{array}{c}
%\begin{widetext}
%\begin{equation}
\displaystyle \left\langle km \left| \frac{e e^*}{|{\bf r}|}
\right|k^\prime n\right\rangle = \frac{2 e e^*}{L} e^{-i \varphi
(m-n)} e^{-i Z(k-k^\prime)}\times
\\ \\
I_{|m-n|} \left(
|k-k^\prime|R\right) K_{|m-n|} \left(
|k-k^\prime|\rho\right)
\end{array}
\label{matrix}
%\end{equation}
%\end{widetext}
\end{eqnarray}
here $I_\mu(x)$ and $K_\mu(x)$ are the modified Bessel functions
\cite{abramovitz} (of imaginary argument) of the order $\mu$,
${\bf r}$ is the vector between the impurity center and the
point on the surface of the nanotube, $R$ is the nanotube
radius. In a cylindrical coordinate system, $\varphi$ is the
angle of the impurity position, $\rho$ is the distance from the
axis of the nanotube to the impurity and $Z$ is its longitudinal
coordinate.

These matrix elements of the potential are needed for calculating
the remote scattering rates. Besides that, the analytical expression for
the remote potential (\ref{matrix}) is interesting by itself. We are
not aware of a calculation of the Coulomb potential for the
charge center {\em removed from the nanotube}. This formula
gives a generalization of an expression for the 1D Fourier
transformation of the Coulomb interaction between charges which
are {\em both on the nanotube surface} (which may be found, for
example, in Ref.\cite{ando-exc}). Let us present analytical
limits of our result at large and small transferred momentum and
demonstrate how the dimension of the nanotube system shows up.

The interaction strength decreases
rapidly with the transferred momentum $q=|k-k^\prime|$, which is
well known property of the Coulomb potential at any dimensions.
Using the asymptote of the Bessel function at $q\gg \rho^{-1},R^{-1}$,
we reduce the Eq.(\ref{matrix}) to
\begin{equation}
\frac{2 e e^*}{L}
e^{-i \varphi (m-n)}
e^{-i Z(k-k^\prime)}
\frac{e^{- |k-k^\prime|(\rho-R)}}{|k-k^\prime|\sqrt{\rho-R}}
\label{matrix2}
\end{equation}
and recover formula $e^{-q a}/q$,
the expression for the Fourier component of
the Coulomb potential in 2D\cite{andofowler}.
This is not surprising, because in
the short--wavelength limit
one restores the planar geometry
when the curvature of the graphite sheet becomes unimportant.

On the other hand, at small $q$ and $m=n$ the matrix element
(\ref{matrix}) logarithmically diverges as $\sim \log (q\rho/2)$,
which is according to 1D electrostatics \cite{loudon}. In
contrast, at $m\neq n$ and small $q$, the limit of the matrix
element has no dependence on $q$. Instead it is proportional to
$(R/\rho)^{|m-m^\prime|}$, and decays exponentially with the
transferred angular momentum $|m-m^\prime|$ as in the multipole
expansion series. It is consistent with understanding of $R^{-1}$
as a minimum cut off momentum in the Coulomb matrix element.

\subsection{Averaging of the random potential
\label{sec:average}}

In the Born approximation each scattering event is statistically
independent and the electron wave function is not coherent
between events. Thus, one has to sum probabilities of the
scattering over the realization of impurities (to be averaged
later). Let us write the partial probability of the single
scattering event using the Fermi golden rule:
\begin{eqnarray}
\begin{array}{l}
%\begin{widetext}
%\begin{equation}
\displaystyle W_{mn}(k)= \frac{2\pi}{\hbar} \sum_q \left(\frac{2
e e^*}{L}\right)^2 I_{|m-n|}^2 \left( qR\right)\times
\\ \\
K_{|m-n|}^2 \left( q\sqrt{h^2+Y^2}\right)
\delta\left( E_{n,k+q}-E_{m,k}\right)
\label{goldenrule}
\end{array}
%\end{equation}
%\end{widetext}
\end{eqnarray}
where $q$ is the transferred momentum, and for the SWNT the sum
has only several terms (less than four) if any, which depends on
$m, n$ and $k$ quantum numbers. Here, $h$ and $Y$ are the
Cartesian coordinates of the impurity (without loss of generality,
we chose the coordinate origin such that $Z=0$). We remind that
the axis of the nanotube is at the distance $X=h$ from the
substrate. $L$ is the tube length.

\begin{figure}[h]
\centerline{
    \includegraphics[width=9cm]{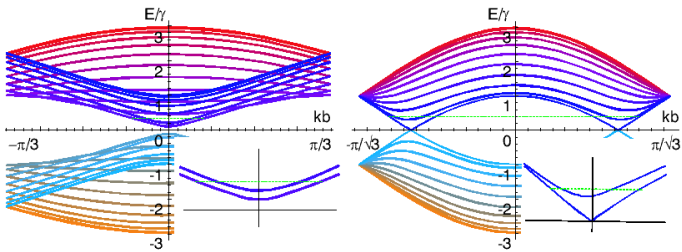}
}
%\vskip .1cm
\caption{\label{fig:spectrum}
Electron energy dispersion for armchair [10,10] and zigzag
[17,0] SWNTs. Inset: Zoom out of the lowest subbands and the
electrochemical potential (green dashed line).} \end{figure}

The Coulomb centers are distributed on the surface of the
substrate randomly. One has to perform averaging in the plane to
obtain a statistical description of the scattering.
We assume that the impurity positions are not correlated. Then
for the electron with the
momentum $k$ in the subband $|m\rangle$,
an elastic lifetime due to the remote impurity scattering is
written as follows:
%%\begin{eqnarray}
%%\begin{array}{c}
%\begin{widetext}
\begin{equation}
\displaystyle \tau^{-1} (m, k)=
\frac{8}{\hbar}
\left(e e^*\right)^2 n_s
\sum_n \sum_{q_k}
\left( \frac{\partial E_{n}}{\partial k}
\right)^{-1}_{k=q_k}
%\right.
%%\\ \\
\frac{{\cal G}(q_k)}{q_k}
%%q_k^{-1}
%%I_{|m-n|}^2 \left( q_kR\right)
%%{\cal F}_{|m-n|}\left(q_kh\right)
%\int^\infty_\infty dy
%K_{|m-n|}^2 \left( q_k\sqrt{h^2+y^2}\right)
%%\end{array}
\label{tau}
\end{equation}
%\end{widetext}
%%\end{eqnarray}
where $n_s$ is a 2D density of the surface impurities and $q_k$
are the solutions of the equation $E_{n,k+q}=E_{m,k}$. As we
noticed before this equation may have up to 4 solutions within the
first Brillouin zone (see Fig.\ref{fig:spectrum}). For example,
for the armchair nanotube we have:
\begin{widetext}
\begin{subequations}
\label{q4}
\begin{equation}
q_k=-k \pm 2 \text{arccos}\left[
-\frac{1}{2} \cos \frac{m\pi}{N} \pm \frac{1}{2}
\sqrt{ \cos^2 \frac{m\pi}{N} +4 \cos \frac{kb}{2}
\left(\cos \frac{kb}{2}+ \cos \frac{n\pi}{N}\right)}\right],
\label{q4a}
\end{equation}
and for the zigzag SWNT:
\begin{equation}
q_k=-k \pm \frac{2}{\sqrt{3}}
\text{arccos}\left[\frac{1}{2\cos\frac{m\pi}{N}}\left(
- \cos \frac{2 m\pi}{N} +  \cos \frac{2n\pi}{N} +
\cos \frac{\sqrt{3}kb}{2} \cos\frac{n\pi}{N}\right)\right].
\label{q4z}
\end{equation}
\end{subequations}

We use the notation ${\cal G}(q_k)=I_{|m-n|}^2 \left( q_kR\right)
{\cal F}_{|m-n|}\left(q_kh\right)$ for a reduced Coulomb matrix
element in Eq.(\ref{tau}). Here the factor ${\cal F}$ comes after
averaging in the plane and equals
\begin{eqnarray}
{\cal F}_{n}(x)&=&\int^\infty_{-\infty} dt
K_n^2 \left(\sqrt{x^2+t^2}\right)=
\frac \pi 2 \int^\infty_{4x} dt K_{2n}(t)= \label{F}
\label{fau}
\\
&=&\frac {\pi^2} 4 \left\{
(-1)^n \left[1-4x
\left(
K_0(4x){\mathbf L}_{-1}(4x)+K_1(4x){\mathbf L}_0(4x)
\right)
\right] +
\frac 4 \pi \sum_{j=1}^n (-1)^{j+1} K_{2(n-j)+1}(4x)
\right\}
\nonumber
\end{eqnarray}
where $\mathbf{L}$ are the modified Struve functions
\cite{abramovitz} (of imaginary argument).

Now we analyze these expressions in the limit of small and large
\(q_k\). For small \(q_kh\) one can write:
\begin{equation}
\displaystyle
q_k^{-1} {\cal G}(q_k)=
q_k^{-1} I_{|m-n|}^2(q_k R)
{\cal F}_{|m-n|}(q_k h)\approx\left\{
    \begin{array}{ll}
       \pi^2/4q_k & \mathrm{if}\quad n=m\\
       \frac{\pi R}{16}\frac{R}{h} & \mathrm{if}\quad |n-m|=1 \\
       \frac{\pi h (2|n-m|-3)!!}{2^{3|m-n|+1}|n-m| |n-m|!} \left(\frac
       Rh\right)^{2|m-n|}& \mathrm{if}\quad |n-m|>1
    \end{array}
\right. \label{Small_q}
\end{equation}
\end{widetext}
The numerical factor is rather small in case of inter--subband
transitions \(n\neq m\) and decreases with the separation
between subbands, \(|m-n|\), rapidly. Thus, the scattering into
the same subband (transition with the orbital quantum number
conservation) is the most effective scattering channel for small
\(q_k\).

For large \(q_kR\) the scattering rate
exponentially decreases with the transferred momentum
$\sim\exp[-2(2h-R)q_k]$, due to the exponential decay of the
modified Bessel function. The main term
of a Poisson series of the matrix element does not depend
on \(|m-n|\):
\begin{widetext}
\begin{equation}
q_k^{-1} {\cal G}(q_k)\approx
\sqrt{\frac\pi{128hR^2q_k^5}}e^{-2(2h-R)q_k} \left(1+\frac{(\frac
hR-\frac14)(|m-n|^2-\frac14)+\frac1{64}}{qh}+\cdots\right)
\label{Large_q}
\end{equation}
\end{widetext}
We will use these analytical expressions (\ref{Small_q}) and
(\ref{Large_q}) for the calculation of the scattering rates in
what follows.

\section{Conductivity: Drift--diffusion approach
\label{sec:dde}}

In this section we calculate the conductivity in a
drift--diffusion (DD) model which is widely used for description
of the transport in multiwall nanotubes. It is also applicable
for very long SWNTs if the phase breaking and/or inelastic
scattering lifetime is short as discussed in Introduction.

It is known, that a Schottky barrier forms near the metal
contact\cite{tersoff,dekker}. For the short channel SWNT device
this Schottky barrier almost determines important transport
characteristics. The theory of the nanotube transport, taking
into account phenomena in the contact regions, is presented in
Refs.\cite{tersoff,dekker} and we do not address this subject in
our paper. Instead we focus on the scattering in the rest of the
tube. The contact region has a finite length which is about a
typical screening length, {\em e.g.}, a distance to the backgate
\cite{odintsov,jetpl}. If this distance is much smaller than the
length of the nanotube, one may define the device channel
conductivity\cite{mceuen-cm,tersoff-vert}. For the sake of
clarity, we restrict ourselves to the case of the armchair or
zigzag SWNT. The generalization of our model to the case of any
chiral SWNT is straightforward.

The conductivity in a single channel is as follows:
\begin{equation}
\sigma_i=\frac{g}{2\pi} \int \frac{e^2 \tau(E)}{\hbar^2}
\frac{\partial E}{\partial k_i} \frac{\partial f}{\partial E}
dE. \label{sigma0}
\end{equation}
Here $g$ is the degeneracy of the current channel. By the
channel of conductivity we understand here any fixed subband of
the orbital quantization which can carry the current. $\tau(E)$
is the transport lifetime; $f(E)$ is an equilibrium distribution
function. The derivative of the distribution function, $\partial
f/\partial E$ is peaked at the electrochemical potential
(delta--function of the Fermi level $E=E_F$ for $T=0$). However,
we keep the integral sign even at $T=0$ because of several
channels corresponding to several non--zero terms in the
conductivity, as given by Eqs.(\ref{q4a},\ref{q4z}). The
$\sigma$ has non--trivial temperature dependence due to the
strong dependence of the lifetime on the electron energy.

It is important to include the inter--subband scattering terms
(if corresponding transition is allowed) because the Coulomb
scattering rate decreases with the transferred momentum (see
also Fig. \ref{fig:abcd}). With increasing $E_F$ the scattering
into the same subband may become less effective than the
scattering into the other subband. We demonstrate below that
this is the case for the semiconductor SWNT at the high doping
levels.

The Eq.(\ref{tau}) can be conveniently rewritten as
\begin{equation}
\displaystyle \tau^{-1} (m, k)=
\frac{8 U_s^2}{\hbar^2}
\mathop{{\sum}^\prime}\limits_{n,q_k}
v^{-1}_{n,q_k} q_k^{-1}
{\cal G}_{|m-n|}\left(q_k\right),
\label{tau2}
\end{equation}
with use of notations: $U_s=\left(e e^*\right)\sqrt{n_s}$ for a
characteristic energy of the Coulomb disorder, and
$v_m=\hbar^{-1}\partial E/\partial k$ for the electron velocity.
The prime sign reminds that the summation is over the roots of
Eqs.(\ref{q4a},\ref{q4z}).

Let us first consider the scattering of the electron in the same
subband $|m,k\rangle \to |m,k+q\rangle$.

%
%\newpage
%

\subsection{Armchair nanotubes: Intra--subband scattering
\label{sec:armch}}

The expression for the scattering roots for the armchair SWNT,
Eq.(\ref{q4a}), in the limit of intra--subband scattering
($m=n$) reduces to:
\begin{equation}
q_k=-k \pm 2 \text{arccos}\left[\cos \frac{m\pi}{N} + \cos
\frac{k}{2} \right], \quad -2k, \quad 0,
\label{qred4a}
\end{equation}
The last root means no scattering because all quantum numbers
are conserved and has to be discarded. The first root is, in
fact, the inter--subband backscattering near the same Fermi
point (see Fig.\ref{fig:spectrum} Right Inset), which is
forbidden for the two crossing subbands of the armchair SWNT by
the pseudospin conservation rule \cite{ando-spin}. The root $q=
-2k$ is the backscattering of the electron within the same
subband to other Fermi point (see Fig.\ref{fig:spectrum} Right).
Only this transition between two Fermi points with a large
transferred momentum ($q_k\approx4\pi/3\sqrt{3}b$ for small
$E_F$) is allowed by the symmetry of the armchair SWNT. Thus,
the analytical limit (\ref{Large_q}) can be used for calculating
$\tau(E)$. The transport lifetime of the electron contains a
large exponential factor
\(\propto\exp\left[\frac{4\pi(2h-R)}{3\sqrt{3}b}\right]\) and,
hence, this mechanism gives the negligible scattering rate.

Because of the exponential dependence of the matrix element on
the transferred momentum, the conductivity of the armchair
nanotube increases dramatically, when the electrochemical
potential reaches the next subband. This subband has a different
symmetry \cite{damnjan} and the inter--subband scattering near
the same Fermi point (see Fig.\ref{fig:spectrum} Right Inset) is
not forbidden. Then, the inter--subband scattering rate of the
armchair nanotube is high and given by an expression which is
similar to the case of zigzag nanotube. We will consider these
two mechanisms together in Sec.\ref{sec:inter}.

\subsection{Zigzag nanotubes: Intra--subband scattering
\label{sec:zigz}}

In the last section we show that the remote impurity scattering
is negligible in the armchair SWNTs. The scattering due to any
other long range potential, except for torsional phonon
modes\cite{kane-mele}, was shown to be ineffective as well
\cite{ando}. The case of the zigzag nanotube is less trivial. We
substitute Eq.(\ref{tau}) into Eq.(\ref{sigma0}) and obtain the
conductivity at zero temperature
\begin{equation}
    \sigma=\frac{e^2 g}{2\pi\hbar \, 8U_s^2}\sum_{m,k_{Fm}}
    \left[\sum_{n,k_{Fn}}
    \frac{\mathcal{G}_{|m-n|}(|\vec
k_{Fm}-\vec k_{Fn})}{v_mv_n|\vec k_{Fm}-\vec k_{Fn}|}
    \right]^{-1},
    \label{sig}
\end{equation}
here $g=4$ is the current channel degeneracy (for the spin and
orbital momentum); $v_m$ is the electron velocity at the Fermi
level (near the $\Gamma$--point), $v_m=\hbar^{-1}\partial
E/\partial k$. Taking into account the non--parabolicity of the
energy dispersion of the $m^{\rm th}$ subband, $E_{k,m}$ we
obtain:
\begin{equation}
    v_m=\frac{6b\gamma}{4 \hbar E_F}\sqrt{E_F^2-E^2_m}=V_F\sqrt{1-\frac{E^2_m}{E_F^2}},
    \label{v}
\end{equation}
here $V_F=3b\gamma/(2\hbar)$ is the Fermi velocity in the
(metallic) massless subband. The Fermi momentum of the electron
in the $m^{\rm th}$ subband is:
\begin{equation}
    k_m(E_F)=\frac2{3b\gamma}\sqrt{E_F^2-E^2_m},
    \label{k}
\end{equation}
where the bottom of the $m^{\rm th}$ subband is:
\begin{equation}
    E_m=\gamma\left|1+2\cos\frac{\pi m}N\right|\simeq \frac{\hbar V_F m}{3R}.
    \label{E}
\end{equation}
These expressions are essentially similar but not equivalent to
ones obtained with $k\cdot p$ method because of
non--parabolicity of the energy dispersion in tight--binding
model.

For the electrochemical potential located within the lowest
subband, $E_m<E_F<E_{m\pm 1}$, the only one level of orbital
quantization is populated at $T=0$, which has
$m=(N+\mathrm{Mod}_3[N])/3$.

Because the inter--subband transitions become allowed only at
$E_F>E_{m\pm 1}$ the single backscattering term with $m=n$ has
to be substituted in the Eq.(\ref{sig}) which gives
\begin{equation}
\sigma_z^{(1)}= G_o g v_m \tau(m,k_{F})= \frac{G_o g\hbar^2}{8
U_s^2}2k_F v_m^2 {\cal G}^{-1}(2k_F) \label{sigm_sm_q}
\end{equation}
here $G_o=e^2/2\pi\hbar$ is the conductance quantum, and the
electron velocity is
\begin{equation} v_m^2(E_F)= V_F^2\times\left\{
    \begin{array}{ll}
        1-\frac{E_m^2}{E_F^2} & \textrm{if}\quad \eta=\mathrm{Mod}_3(N)\ne0 \\
      1  & \textrm{if}\quad \eta=\mathrm{Mod}_3(N)=0
    \end{array}
\right.
\label{vm_q}
\end{equation}
here the index $\eta=\mathrm{Mod}_3(N)$ distinguishes between
the zigzag metallic ($\eta=\mathrm{Mod}_3(N)=0$) and
semiconductor SWNTs ($\eta=\mathrm{Mod}_3(N)\ne0$).

Let us now apply the expression (\ref{Small_q}) for analysis of
the conductivity at small $E_F$. Because at small $k_F$ the
function ${\cal G}^{-1}(2k_F)$ does not depend on $k_F$ in the
leading term, we find the DD conductivity of the zigzag tube at
the low doping level depends on the concentration as:
\begin{equation}
\sigma_z^{(1)}=\frac{G_o g E_F \hbar V_F}{4 U_s^2}
{\cal G}^{-1} \times \left\{
    \begin{array}{ll}
      \left(1-\frac{E_m^2}{E_F^2} \right)^{3/2} &
                \textrm{if}\quad \eta\ne0, \\
      1 & \textrm{if}\quad \eta=0.
    \end{array}
\right.
\label{sigma_z_sm}
\end{equation}
We drop the argument of ${\cal G}^{-1}$ in the expression above
because its limit is $4/\pi^2$ at small $k_F$. The DD conductivity
is different for the metallic and semiconductor tubes: it is
linear in $E_F$ for the metallic (zigzag, $\eta=0$) SWNT and it
depends on the Fermi energy as
$$E_F \left( 1-\frac{E_m^2}{E_F^2}\right)^{3/2}$$
for the degenerately doped semiconductor SWNT (zigzag,
$\eta\ne0$).

The conductivity at small $E_F$ depends linearly on $E_F$ owing
to the matrix element of the Coulomb potential of the 2D remote
centers. The square of the matrix element is $\sim q^{-1}$,
because in our model the centers are distributed on the (2D)
surface of the substrate (the generalization to the 3D case is
obvious). The linear dependence of the square of the matrix
element in $q^{-1}$ results in the linear dependence of the
conductivity on $E_F$.

\begin{figure}[h]
\centerline{
    \includegraphics[width=3in]{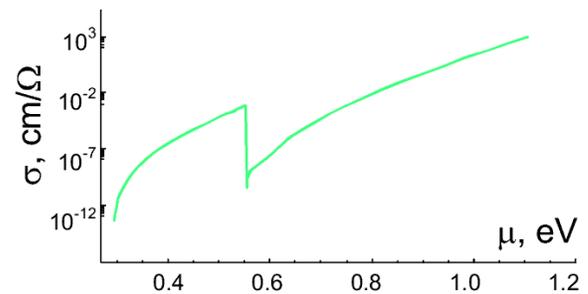}
}
\caption{
    \label{fig:sigma}
Conductivity of a zigzag [17,0] SWNT vs. the electrochemical
potential (the doping level). } \end{figure}

At larger $E_F$ the energy dependence of the transport lifetime
(of the matrix element of the transition) is different. In this
case, the Eq.(\ref{Large_q}) has to be used, which results in a
fast exponential growth of the conductivity because of the large
suppression of the transitions with increasing momentum transfer.
Then, the Eq.(\ref{sigma_z_sm}) transforms into:
\begin{widetext}
\begin{eqnarray}
\begin{array}{c}
\displaystyle \sigma_z^{(1)}
\simeq \frac{8 R G_o g E_F^2}{U_s^2}
\sqrt{\frac{h E_F}{\pi\hbar V_F} \sqrt{1-\frac{E_m^2}{E_F^2}}}
\left(1-\frac{E_m^2}{E_F^2}\right)^2
\exp\left[4(2h-R)\frac{E_F}{\hbar V_F}
\sqrt{1-\frac{E_m^2}{E_F^2}}\right]
\\ \\
\displaystyle \stackrel{\epsilon_F\to 1}{%\longrightarrow
\sim} \frac{2^{\frac{21}{4}} R G_o g E_F^2}{U_s^2} \sqrt{\frac{h
m}{3\pi R}} \left(\epsilon_F-1\right)^{9/4} \exp\left[8(2h-R)
(\epsilon_F-1)\right]
\end{array}
\label{sigm_z_g}
\end{eqnarray}
\end{widetext}
here $\epsilon_F=E_F/E_m$ is the dimensionless Fermi level of
the semiconductor tube: $\epsilon_F \to 1$ and $\epsilon_F<1$ in
the expression (\ref{sigm_z_g}). This equation gives the
conductivity of the zizgaz nanotube for the transport through
only one channel (of the lowest subband). The result of our
calculation (the Eqs.(\ref{sigma_z_sm}) and (\ref{sigm_z_g})) is
presented in the Figure \ref{fig:sigma}, where the
drift--diffusion conductivity of the zigzag SWNT is plotted as a
function of the Fermi level (doping level). With increasing
doping level the conductivity grows exponentially. This is not
because of more carriers are available for transport but due to
the energy dependent transport lifetime that grows exponentially
with the increase of the momentum transfer between initial and
final states. This momentum transfer is $2k_F\propto E_F$ when
the linearization of the TB Hamiltonian is possible. The
momentum transfer increases until the Fermi level reaches the
next subband. Then, suddenly, the new backscattering channel
opens. The momentum transfer between the next subband and the
lowest subband is small at this critical doping level. Thus, the
remote impurity scattering becomes very efficient and the
conductivity drops several orders of magnitude.

%
%\newpage
%

The maximum conductivity may be reached when the Fermi level is
close to the edge of the second subband but lower than it. We
note that the phonon scattering may limit the transport, at
least, at high temperatures \cite{phonon}. Therefore, for not
very low temperature and high enough doping level, it may be
possible to observe switching from the remote scattering to the
phonon scattering mechanism.

Also at $T\neq 0$, the finite temperature distribution function
makes the $\sigma(E_F)$ dependence smooth near the step at
$\epsilon_F\sim 1$, as it will be shown next.

\subsection{Inter--subband scattering  \label{sec:inter}}

\begin{figure}[h]
\centerline{
    \includegraphics[width=3.5in]{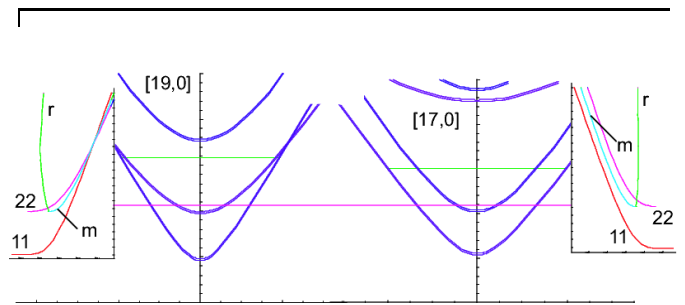}
} \caption{
    \label{fig:abcd}
Bandstructures for zigzag SWNTs of two types. Left: [3q+1,0]
SWNT; Right: [3q-1,0] SWNT. Doping level is shown as horizontal
lines: pink/green (lower/upper) line is for low/high doping
level. Insets show the scattering rates for different scattering
mechanisms as a function of energy as in left and right diagrams
respectively.}
\end{figure}

The remote impurity scattering between the subbands may happen
only when the doping level is high enough to essentially
populate the second subband, $E_F\ge E_{m\pm 1}$. Then, the
scattering rate becomes high and the mean free path becomes
short. The expressions for the DD conductivity of armchair
metallic, zigzag quasi--metallic and zigzag semiconductor
nanotubes are essentially the same in this region. This is
because the DoS in the vicinity of the Fermi point is a
universal function\cite{minitm} of the energy (doping level),
and because there is no special selection rule for the
transitions {\em between different subbands} of the orbital
quantization. Let us consider the single Fermi point and find
which scattering channel defines the conductivity at $E_F\ge
E_{m\pm 1}$. As before, because of the large momentum transfer,
we neglect transitions between different Fermi points, which are
possible for the armchair SWNT. As shown in Fig.\ref{fig:abcd},
we have two left--going and two right--going (current)
states/channels (to be multiplied with the spin and orbital
momentum degeneracy $g=4$). We introduce four scattering rates
$\tau_{ij}^{-1}$ and calculate it with the Eq.(\ref{tau2}).

Let us assign the index $i=1$ to the subband with the largest
momentum at the Fermi level, $k_{Fi}$ (which may change with
changing the Fermi level if the subbands cross). The
intra--subband lifetimes $\tau_{11/22}$ are given by terms of
Eq.(\ref{tau}) with $n=m$. There are two inter--subband
lifetimes: $\tau_r^{(12)}$ and $\tau_m^{(12)}$ for the
transition with the same/opposite sign of the electron velocity
in the initial and final states $|n>$ and $|m>=|n\pm 1>$.

For certain tube symmetry the subband crossing may happen with
increasing $E_F$. When the Fermi energy is above the crossing
point the subbands 1 and 2 are interchanged in the equations
given below. The case of the metallic (armchair or zigzag) SWNT
is similar to the case of noncrossing subbands and will not be
considered separately. There is only one difference for the
armchair SWNT as compared with the zigzag SWNT case: the
dispersion in the subband 1 is massless. Thus, $v_1=V_F$ is the
constant and, by symmetry, $\tau_{11}=\infty$ ("intra--subband"
scattering in the lowest subband is not allowed).

We recalculate the distribution functions in all 4 channels
using Boltzmann equation and taking into account the
inter--channel transitions. When the scattering is weak one can
neglect it and use the equilibrium distribution function. This
is not the case for the SWNT at $\epsilon_F\ge 1$, where the
scattering rates at the edge of the second subband are very
high.

General expression for the DD conductivity with the non--zero
inter--subband scattering is rather cumbersome even in the
approximation of two closest subbands:
\begin{widetext}
\begin{eqnarray}
\begin{array}{l}
\sigma =G_o g \frac{ \displaystyle v_1 \tau_r
+\frac{(v_1+v_2)^2}{2v_1}\tau_{22}
-\frac{(v_1-v_2)^2}{2v_1}\frac{\tau_r}{\tau_m}\tau_{22}
+v_2\frac{\tau_r}{\tau_{11}}\tau_{22}} { \displaystyle
1+\frac{\tau_r}{\tau_m}+\frac{2v_2}{v_1}\frac{\tau_{22}}{\tau_m}
+\frac{2\tau_r}{\tau_{11}}
+\frac{v_2}{v_1}\frac{\tau_{22}}{\tau_{11}}\left(1
+\frac{\tau_r}{\tau_m}\right)}
%\\ \\
\simeq {\displaystyle \frac{G_o g \hbar^2}{8 U_s^2}} \left.[ v_1
v_2 |k_1-k_2| {\cal G}^{-1}(|k_1-k_2|) \left(1- \right. \right.
\\ \\
\left. \left. \displaystyle
 \frac{|k_1-k_2| {\cal
G}^{-1}(|k_1-k_2|)}{ |k_1+k_2| {\cal G}^{-1}(|k_1+k_2|)} \right)
%\right.
%\\ \\
%\left.
\displaystyle +\frac{(v_1+ v_2)^2v_2}{2v_1} 2k_2 {\cal
G}^{-1}(2k_2)
\left(1-\left(1+\frac{(v_1-v_2)^2}{(v_1+v_2)^2}\right)
\frac{|k_1-k_2| {\cal G}^{-1}(|k_1-k_2|)}{ |k_1+k_2| {\cal
G}^{-1}(|k_1+k_2|)} \right) +\dots \right],
\end{array}
\label{intersub}
\end{eqnarray}
\end{widetext}
and we study below the limiting cases where simpler analytical expressions are available.

At small $k_{F2}$, at the second subband edge (see
Fig.\ref{fig:abcd}, the Fermi level is at lower/pink line), the
momentum transfer of the inter--subband scattering
$(1\rightleftarrows 2)$ is larger than of the intra--subband
scattering $(2\rightleftarrows 2)$ (see Figure \ref{fig:abcd}
Inset) and the last scattering channel is more effective. The
corresponding contribution to the $\sigma$ is $\propto v_2
\tau_{22}$, where the velocity at the Fermi level is given by
Eq.(\ref{v}) (the lowest subband $i=1$ has the larger velocity).

According to inequality:
\begin{equation}
\tau^{-1}_{r}>\tau^{-1}_{22}>\tau^{-1}_{m}>\tau^{-1}_{11}
%\label{ineqrates}
\end{equation}
the most important term in the $\sigma$ is $\propto
v_1\tau_{r}$, if the shortest time is $\tau_{22}$. Thus the DD
conductivity reads as
\begin{equation}
\sigma=G_o g v_1 \tau_{r}+\dots \simeq \frac{G_o g \hbar^2}{8 U_s^2} v_1 v_2 |k_1-k_2|\, {\cal
G}^{-1}(|k_1-k_2|), \label{mainterm}
\end{equation}
where $v_1=V_F\sqrt{1-(\hbar V_F)^2/(3R E_F)^2}$ and
 $v_2=V_F\sqrt{1-(2\hbar V_F)^2/(3R E_F)^2}$.

%
%\newpage
%

With increasing $E_F$ (see Fig.\ref{fig:abcd}, Fermi level at
upper/green line), since $q_{22}>q_{12},$ then,
$\tau_{22}<\tau_{r}$ (see Inset of Figure \ref{fig:abcd}), and
the leading term of Eq.(\ref{intersub}) is due to the
intra--subband transition ($2\rightleftarrows 2$). Then, for the
armchair and zigzag ($\eta\neq +1$) nanotubes, the main term in
the DD conductivity is
\begin{eqnarray}
\begin{array}{l}
\displaystyle \sigma=G_o g \frac{(v_1+ v_2)^2}{2v_1} \tau_{22}+\dots
\\ \\
\displaystyle \simeq \frac{G_o g \hbar^2}{8 U_s^2}\frac{(v_1+ v_2)^2 v_2}{2v_1} 2k_2\, {\cal G}^{-1}(2k_2).
\end{array}
\label{mainterm2}
\end{eqnarray}

In the zigzag $\eta= +1$ SWNT, the crossing of the lowest
subbands occurs in the studied region (shown in Fig.
\ref{fig:abcd} Left). In that case $k_{F1}>k_{F2}$ for the $E_F$
is lower than the crossing point, and $k_{F1}<k_{F2}$ for the
$E_F$ is higher than the crossing point. So the greatest
scattering rate is due to ($2\rightleftarrows 2$) intra--subband
transition below the crossing point and ($1\rightleftarrows 1$)
intra--subband transition above the crossing point (see Left
Inset of Fig. \ref{fig:abcd}). Thus, the indices 1 and 2 must be
exchanged in the Eq.(\ref{intersub}) and Eq.(\ref{mainterm2}).

By considering these two cases we cover all possible situations and present possible analytical expressions for
the DD conductivity within the remote center scattering model.

\section{Conductivity at the finite temperature and impurity potential fluctuation
\label{sec:thresh}}

In the last section we studied the conductivity of the nanotube
in the zero temperature limit. The temperature dependence adds
to the above result via substituting Fermi--Dirac distribution
functions in Eq.(\ref{sigma0}) instead of step functions as we
implicitly used before.

\begin{figure}[h]
\centerline{
    \includegraphics[width=3in]{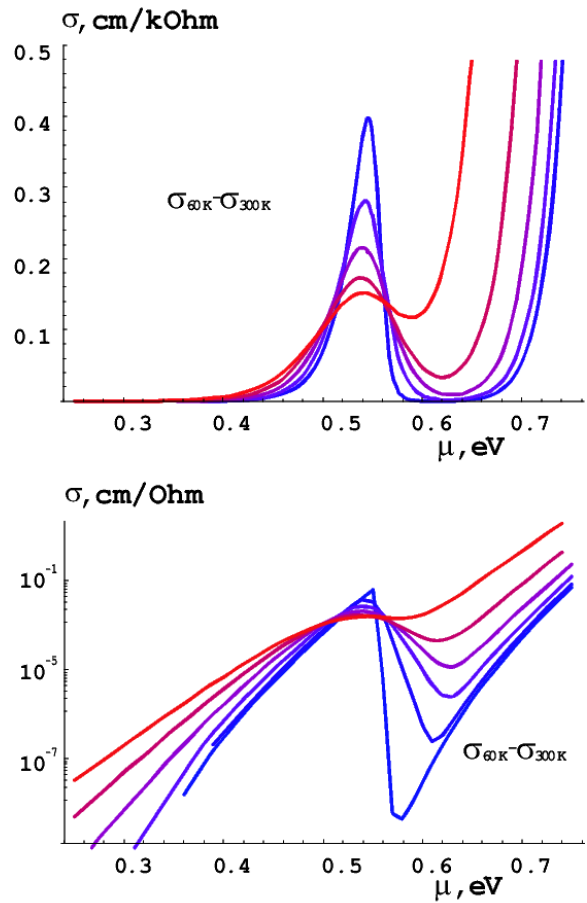}
} \caption{
    \label{fig:temp}
Temperature dependence of the conductivity of a zigzag [17,0]
SWNT vs. the electrochemical potential in a vicinity of the
second subband edge. }
\end{figure}

We present the numerical result on the temperature dependence of
the conductivity in Fig.\ref{fig:temp}.

%
%\newpage
%

We already discussed an important assumption of our model: the
phase breaking time has to be short enough which seems to
fulfill for the nanotubes due to the e--e interaction and
scattering of the electron into the channels/bands which are
different from the transport channel/band. This allows one to
neglect the interference correction which is normally dominating
in pure 1D systems. However, the bare 1D Coulomb potential may
still localize the carriers in a nanowire for the infinite
channel length. We remind that the {\em quasi--one--dimensional}
potential created at the nanotube channel by the random
distribution of remote impurities on the substrate surface has
the cut--off length and, thus, a maximum amplitude of the
potential. The localization length in a finite system is defined
by the average fluctuation of the random potential. Our theory
is applicable only in the limit of the electrochemical potential
which is much higher than the average fluctuation of the random
potential of the remote scatterers.

Since the Coulomb centers are located on the substrate and the
1D electron is on the nanotube, there are two different types of
averaging for the 2D distribution of impurities and for the 1D
random potential for the electron. The operator of the Coulomb
potential is given by the Eq.(\ref{potential}). It creates the
1D potential along the nanotube, which reads as:
\begin{equation}
U_{1D}(z) = \sum_i V(Y_i,Z_i),
\label{1d}
\end{equation}
where the sum is over impurities that have random positions.

To calculate the average fluctuation of this potential along the
nanotube we average it over the impurity positions:
\begin{equation}
\delta U=\sqrt{\langle U^2_{1D}\rangle -\langle U_{1D}\rangle^2}=
2\sqrt{2+2\log 2+\log^2 2}\; e e^* \sqrt{n_s}.
\label{fluct}
\end{equation}
At low temperature, for random impurity distribution, this
average fluctuation gives an estimate for a lower bound of the
electrochemical potential at which the Born approximation for
the scattering is applicable:
\begin{equation}
E_F\gg U_s
\label{born}
\end{equation}

It is known that the condition of applicability of the Boltzmann
equation is that the Fermi energy must be much larger than the
inverse scattering time. If we estimate the latter as
$\hbar/\tau\sim4U_s^2/E_F$ then we arrive to the condition which
is similar to Eq.(\ref{born}).

\section{Effective impurity charge and SWNT depolarization
\label{sec:depol}}

Though, the actual charge of impurity is not known, we assume it
to be an elementary charge $e$. However, the substrate
polarization results in a reduction of this value. As long as
the distance between the charge center and the substrate surface
is much smaller than all other lengths of the problem: $R$, $h$,
etc., one may use an effective dielectric function of the
substrate to define the effective charge as $e^*\to
2e/(\varepsilon +1)$ where $\varepsilon$ is the dielectric
function of the substrate and unity stays for the permittivity
of the vacuum. It is the leading term of expansion series of the
image charge potential, which has to be kept in the remote
scattering calculation.

One must take into consideration an effect of depolarization of
the Coulomb potential due to the screening by carriers in the
nanotube. This changes the remote scattering potential
essentially. We calculate this effect using a continuum model
for SWNT electrostatics \cite{jetpl,ecs}. Within the model, the
depolarization of the SWNT at the distance $D_g$ from a
conducting gate is given by the following expression:
\begin{equation}
V_{eff}=\displaystyle \frac{1}{1+\alpha} V =
\frac{C_Q^{-1}}{C_g^{-1}+C_Q^{-1}} V,
\label{depolariz}
\end{equation}
where the depolarization $\alpha$ is written in terms of the
geometric capacitance per unit length of the SWNT $C_g^{-1}=2
\log (2D_g/R)$ and the quantum capacitance per unit length,
which equals $C_Q^{-1}=1/(e^2 \nu_M)$ for the metallic and
degenerately doped semiconductor tube.

The depolarization can be taken into account together with the
substrate image charge effect simultaneously:
\begin{equation}
e^*= \displaystyle e\frac{2}{\varepsilon +1}
\frac{C_Q^{-1}}{C_g^{-1}+C_Q^{-1}}.
\label{effective}
\end{equation}

\section{Conclusion}

In summary, we have developed a microscopic quantum mechanical
model of an electron scattering by remote Coulomb impurities
lying on the substrate surface. We consider a uniform 2D
distribution of the charged impurities. Theory is readily
generalized for the three--dimensional distribution of the
impurities which case is applicable to the modeling of an
insulator surrounding the nanotube channel (to be published
elsewhere). A general expression for a Fourier transform of the
Coulomb potential of a charge which is removed from the nanotube
surface is given. The matrix element for calculating the remote
impurity scattering in the nanotubes is derived. We obtain a
zero and finite temperature conductivity within a
drift--diffusion transport model. We assume that the e--ph and
e--e scattering between the electrons in the $\pi$--electron
band (transport band) and the electrons in other bands (or/and
other tubes in the rope, or/and other walls in a multiwall
nanotube) presents a fast mechanism of the phase breaking. It
allows us to use the Boltzmann equation and neglect interference
corrections for the classical conductivity. Selfconsistent
calculation of the SWNT depolarization factor, taking into
account image charges in the substrate, is performed and yields
an effective charge of the impurity, used for computing the
scattering rate.

The remote impurity scattering is almost negligible for the
armchair SWNTs if the Fermi level is below the edge of the
second subband, which is consistent with other calculations for
other scattering mechanisms with long range potentials. In
contrast, the scattering rate for the zigzag SWNT is high
enough. In general, the DD conductivity of the SWNT is $\sim G_o
\Lambda$, where the mean free path, $\Lambda\sim v\tau$.
Analytical expressions for the conductivity are obtained in the
limit of small and large momentum transfer. We studied
numerically dependence of the conductivity on the Fermi level
position and found that the highest conductivity (of a
semiconductor SWNT) may be observed for the degenerately doped
SWNT when the Fermi level is close to (but lower than) the
second subband edge. The dependence $\sigma(E_F)$ is not
monotonic. At even higher doping level, the conductivity is low
when we take into account the inter--subband scattering. This is
because $\Lambda$ becomes short with opening a new scattering
channel, when the Fermi level is higher than the second subband
edge. Thus, over--doping of a semiconducting nanotube does not
improve its transport properties. Our theory may be applicable
for multiwall nanotubes, although, an additional analysis will
be required, especially because of different screening.

%\section*{Acknowledgments}
{\bf Acknowledgments.} S.V.R.~acknowledges DoE support through
grant DE-FG02-01ER45932, ONR through grant NO0014-98-1-0604, ARO
through grant DAAG55-09-1-0306 and NSF support through grant No.
9809520. Authors are grateful to Professor K. Hess, Professor U.
Ravaioli for valuable discussions and especially to Professor
A.L. Efros for pointing on mechanisms leading to electron phase
breaking and delocalization in a nanotube quasi--1D system.

%\end{references}


\begin{thebibliography}{99}
\bibitem{ftnt}
Though, an external potential which breaks mirror reflection
symmetry of the nanotube may result in very effective
scattering.
\bibitem{footnote1}
The scattering in the nanotube channel is one of the contributions
to the total device impedance. Another scattering mechanism is a
reflection at the (tunnelling) contact. We do not consider the
contact effects in the paper as it had been treated extensively in
recent publications \cite{tersoff,dekker}.
\bibitem{footnote2}
We note that the SWNT itself is likely free of substitutional
impurities for a high mobility device. Though, the
generalization of our theory for the case of substitutional
impurities is also possible.
\end{thebibliography}
\end{document}